\documentclass[a4paper,12pt]{article}

\usepackage[utf8]{inputenc}
\usepackage{authblk}
\usepackage{hyperref}
\usepackage{graphicx}
\usepackage{geometry}
\usepackage{breakcites}
\usepackage{graphicx}
\usepackage[round]{natbib}

\title{Modeling the molecular impact of SARS-CoV-2 infection on the renin-angiotensin system}

\author{Fabrizio Pucci, Philippe Bogaerts, Marianne Rooman\footnote{To whom the correspondence should be addressed: mrooman@ulb.ac.be}}

\affil{Computational Biology and Bioinformatics,\\ Universit\'e Libre de Bruxelles, CP 165/61, Roosevelt Ave. 50, 1050 Brussels, Belgium} 
\begin{document}

\maketitle

\begin{abstract}
    SARS-CoV-2 coronavirus infection is mediated by the binding of its spike protein to the angiotensin-converting enzyme 2 (ACE2), which plays a pivotal role in the renin-angiotensin system (RAS). The study of RAS dysregulation due to SARS-CoV-2 infection is fundamentally important for a better understanding of the pathogenic mechanisms and risk factors associated with COVID-19 coronavirus disease, and to design effective therapeutic strategies. In this context, we developed a mathematical model of RAS based on data regarding protein and peptide concentrations; the model was tested on clinical data from healthy normotensive   and hypertensive individuals. We then used our model to analyze the impact of SARS-CoV-2 infection on RAS, which we modeled through a down-regulation of ACE2 as a function of viral load. We also used it to predict the effect of RAS-targeting drugs, such as RAS-blockers, human recombinant ACE2, and angiotensin 1-7 peptide, on  COVID-19 patients; the model predicted an improvement of  the clinical outcome for some drugs  and a worsening for others.
\end{abstract}

\section*{Introduction}

Since December 2019, the world has been facing a global viral pandemic of the novel severe acute respiratory syndrome coronavirus 2, ‘SARS-CoV-2’; this pandemic has, to date, caused millions of people to be infected and hundreds of thousands to die \citep{JHU}. First detected in the city of Wuhan (China) \citep{WU2,WU3,WU4,WU5}, SARS-CoV-2 spreads rapidly throughout the world. The coronavirus family, to which SARS-CoV-2 belongs, includes a number of viruses, such as SARS-CoV and MERS-CoV, that have been implicated in serious epidemics that cause acute respiratory distress syndrome (ARDS). There is not yet consensus on the origin of the SARS-CoV-2   \citep{Origin0,Origin1,Origin2,ACE2}, but the primary hypothesis is that it originated from bat (\emph{Rhinolophus affinisor}) or pangolin (\emph{Manis javanica}), since the genomes of these two viral species share high sequence identity with SARS-CoV-2.

Coronaviral genomes encode a series of structural proteins, one of which is the spike glycoprotein or S-protein that protrudes from the membrane surface \citep{ACE2}. 
Similar to the SARS-CoV virus that was identified in 2003, the S-protein of SARS-CoV-2 has been shown to bind to the angiotensin-converting enzyme 2 (\texttt{ACE2}) so that it can be used as an entry receptor to the cell \citep{ACE2,ACE2_0,ACE2_1,ACE2_2,ACE2_3}. This protein plays a pivotal role in the renin-angiotensin system (RAS) signaling pathway  \citep{ANG} by  cleaving  angiotensin I and II peptides to generate angiotensin 1–9 and the  biologically active peptide angiotensin 1–7, respectively  \citep{ANGI,ANGII}. 
\texttt{ACE2} is highly expressed in type II alveolar cells of lung, epithelial cells of oral mucosa, colon enterocytes, myocardial cells and kidney proximal tubule cells. 
The protective role of ACE2 in severe ARDS is also widely recognized  \citep{Imai1,Imai2}. Indeed, it has been shown, both \emph{in vitro} and \emph{in vivo} mouse models, that a loss of \texttt{ACE2} expression causes increased production of angiotensin II, and that this contributes to lung failure \citep{Imai2}

It has already been established years ago that the SARS-CoV spike protein interferes with  RAS  due to its binding to ACE2 \citep{WEN}, thus causing ACE2 downregulation; this has opened up a number of interesting means of tackling SARS-CoV infection through RAS modulation. Indeed, injection of a soluble form of recombinant human \texttt{ACE2} (\texttt{rhACE2}, GSK2586881) into mice infected with SARS-CoV appears to have a double role \citep{Imai2}:  it slows the viral infection by binding to the S-protein and rescues \texttt{ACE2} activity, thus causing angiotensin II reduction and protecting lung from severe failure.
 
 \texttt{rhACE2} has been tested in phase II trials for its ability to ameliorate ARDS \citep{TRIAL1}. Although \texttt{rhACE2} treatment is well tolerated by patients and it offers a significant reduction in angiotensin II level, the clinical distress severity was not reduced in a recent pilot study \citep{TRIAL1}. Further studies are needed to understand the biological differences between the responses of animal models and humans. 
 
 Since SARS-CoV-2 also targets  \texttt{ACE2} receptors when it infects cells, it is logical to hypothesize that \texttt{rhACE2} might help reduce the severity of COVID-19 disease \citep{NEWACE}. Indeed, it has been shown that   \texttt{rhACE2} inhibits SARS-CoV-2 infection \emph{in vitro}, and that this inhibition depends both on the initial quantity of the virus and on \texttt{rhACE2} concentration \citep{Vanessa}. Following these exciting results, a clinical trial with exogenous submission of   \texttt{rhACE2} recently started  \citep{TRIAL2}. A number of other clinical trials are also underway that target the dysregulated RAS system to restore its functionality \citep{TRIAL3,TRIAL4,TRIAL5,TRIAL6,TRIAL7}.

Hypertension and cardiovascular disease have been shown to be risk factors in cases of SARS-CoV-2 infection. This brings into question what might be the potential effects on COVID-19 development of the RAS-targeting drugs that are used to treat hypertension and cardiovascular disease. 
RAS-targeting drugs fall into three categories: (i) angiotensin converting enzyme inhibitors (ACE-I), (ii) angiotensin receptor blockers (ARB), and (iii) direct renin inhibitors (DRI) (Fig. \ref{2a}). Several recent studies on large patient cohorts \citep{BIGIF01,BIGIF02,BIGIF03} conclude that there is only weak correlation between treatment with drugs from these categories and any substantial increase in risk of COVID-19 disease. 

Despite these interesting findings, there is not yet a detailed understanding of how SARS-CoV-2 infection leads to a dysregulation of  RAS  and, in severe cases, to ARDS. It is of fundamental importance that we gain better insights into the perturbed RAS  in order to properly elucidate the pathogenic mechanisms and associated risk factors of SARS-CoV-2 infection; this, in turn, will enable novel therapeutic strategies to be designed and tested so that disease progression can be inhibited.


\section*{Results}

The main objective  of this paper is to investigate the effect of RAS-targeting drugs and SARS-CoV-2 infection, both individually and in combination, on RAS. We started by setting up the dynamical model describing RAS of healthy normotensive and hypertensive individuals. The robustness and predictive power of our model was  assessed by investigating the effects of three types of antihypertensive drugs: (i) ACE-I, (ii) ARB and (iii) DRI. This assessment included a comparison of model simulations with patient clinical data.  
Following confirmation of model robustness and accuracy, ACE2 downregulation due to viral infection was introduced into the model to quantitatively predict how RAS is perturbed in COVID-19.

\subsection*{Modeling the renin-angiotensin system}

The RAS system has been widely studied  \citep{REVIEWRAS1,Reninbook1,Reninbook2}. It  plays a key role in the regulation of a large series of physiological systems among which the renal,  lung and  cardiovascular systems. Consequently, its dysregulation is related to multiple pathological conditions such as  hypertension and  ARDS, just to mention some of them
\citep{REVIEWRAS2,REVIEWRAS3,REVIEWRAS4,REVIEWRAS5,BIGREV}. 

\begin{figure}[h!]
    \centering
    \includegraphics[width=0.97\linewidth]{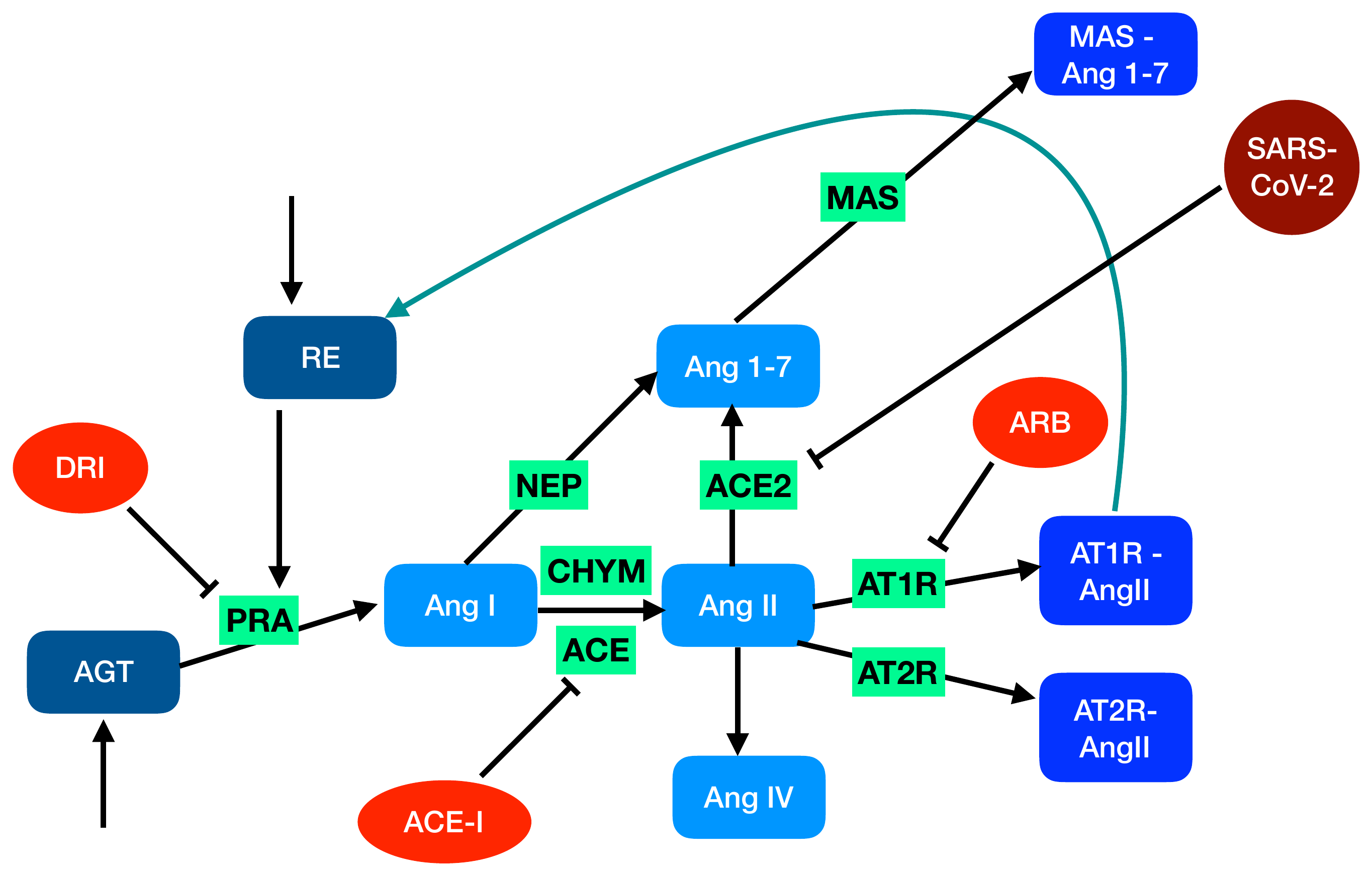}
    \caption{Schematic representation of the  RAS system. In the unperturbed  system,  soluble proteins that are explicitly considered in the model are in grey blue, the peptides in light blue and the peptide-bound membrane proteins  in mid blue. The activities and enzymes considered only through reaction rates are in green. The feedback loop is indicated in blue. In the perturbed  system, the drugs are in orange and SARS-CoV-2 in dark red.  }
    \label{2a}
\end{figure}

There are two different types of RAS: the circulating RAS that is localized in the plasma and is involved in the regulation of the cardiovascular system, and the tissue-localized systems that act intracellularly or interstitially within different organs in association with the systemic RAS or independently of it. Here we focus on the local RAS within the pulmonary circulation and model its network of biochemical reactions schematically depicted in Fig. \ref{2a}.

When the blood pressure decreases, the juxtaglomerular kidney cells that sense changes in renal perfusion pressure secret an aspartic protease protein called renin (\texttt{RE}, EC 3.4.23.15). The activity of this enzyme, called plasma renin activity ($PRA$), is the common measure used in clinical practice to set up diagnosis and treatment design of essential hypertension. 

The dynamics of the renin concentration can be modeled as: 
\begin{eqnarray}
 \frac{d[\texttt{RE}]}{dt} = \beta  -\frac{\texttt{Log}\; 2}{h_{re}} [\texttt{RE}]
 \label{REN}
 \end{eqnarray}
\noindent
where $h_{re}$ is renin's half-life and $\beta$ its production rate. The latter is not constant but depends on other elements of the RAS system which we will discuss  later in the section. 
The role of  renin is to cleave the N-terminus of a protein from the serine protease inhibitor family called angiotensinogen (\texttt{AGT}) to form the decapeptide hormone angiotensin I (\texttt{AngI}). Assuming non-linear Michaelis-Menten kinetics,  the dynamics of the angiotensinogen can be written  as:

\begin{equation}
 \frac{d[\texttt{AGT}]}{dt} = k_{agt} - k^{re}_{cat} \frac{[\texttt{RE}][\texttt{AGT}]}{[\texttt{AGT}]+ K^{re}_M} -\frac{\texttt{Log}\; 2}{h_{agt}} [\texttt{AGT}]
     \end{equation}
\noindent
where $k_{agt}$ is \texttt{AGT}'s production rate, $h_{agt}$ its half-life, $k^{re}_{cat}$ the turnover number of the enzymatic reaction and $K^{re}_{M}$ the  Michaelis constant. Although
the substrate concentration $[\texttt{AGT}] \sim K^{re}_M$ and   thus influences the reaction rate, the \texttt{AGT} concentration is much larger than the \texttt{RE} concentration which, 
as a consequence, impacts more on  RAS regulation. Eq. (2) can thus be linearly approximated as:

\begin{equation}
 \frac{d[\texttt{AGT}]}{dt} = k_{agt} - c_{re}[\texttt{RE}] -\frac{\texttt{Log}\; 2}{h_{agt}} [\texttt{AGT}]
 \label{AGT}
     \end{equation}
\noindent
where the reaction rate $c_{re}$  relates the renin concentration to its activity.

 The \texttt{AngI} peptide is further cleaved by different enzymes: 
 
  \begin{itemize}
  
   \item The angiotensin-converting enzyme (\texttt{ACE}, EC3.4.15.1),  a zinc metalloproteinase located mainly in the capillaries of the lungs and in the endothelial cells. It catalyzes the transformation of \texttt{AngI} into the octapeptide angiotensin II (\texttt{AngII}).
   
   \item Chymase (\texttt{CHY}, EC 3.4.21.39), a serine protease that is mainly localized in blood vessels and heart. It also catalyzes the transformation of \texttt{AngI} into \texttt{AngII}. 
      
       \item Neprilysin (\texttt{NEP}, EC3.4.24.11), another  zinc metalloproteinase that is expressed in a  wide variety of tissues. It catalyzes the transformation of \texttt{AngI} into the heptapeptide hormone angiotensin-(1-7) (\texttt{Ang1-7}).
  \end{itemize}
 
 \noindent
The dynamics of \texttt{AngI} can thus be described as: 

\begin{eqnarray}
 \frac{d[\texttt{AngI}]}{dt} =   c_{re} [\texttt{RE}] - \left( c_{ace} + c_{chy}  + c_{nep}\right) [\texttt{AngI}] -\frac{\texttt{Log}\; 2}{h_{angI}} [\texttt{AngI}]
    \end{eqnarray}
\noindent
where $c_{ace}$, $c_{chy}$ and $c_{nep}$ are the reaction rates associated with the corresponding enzymatic reactions. To get this relation, we started from the non-linear Michaelis-Menten kinetic term, which reads for \texttt{ACE}: ${[\texttt{ACE}][\texttt{AngI}]}/{([\texttt{AngI}]+ K^{ace}_M})$. As the substrate concentration [AngI] is here much lower than the Michaelis constant  of the reaction ($[\texttt{AngI}]<<K^{ace}_M$), we dropped it from the denominator and considered  the equilibrium concentrations of the \texttt{ACE} enzyme fixed, so that the reaction term becomes linear in the concentration of the \texttt{AngI} substrate. We made the same approximation for the reactions involving \texttt{CHY} and \texttt{NEP} and for all other reactions described below.

The role of \texttt{AngII}  in  RAS  is central since it has a vasoconstriction effect, enhances  blood pressure, and triggers inflammatory processes and fibrosis. In lung, the capillary blood vessels are among the sites that have the highest \texttt{ACE} expression and production of \texttt{AngII}. 
Its dysregulation has frequently been  related to a wide series of chronic and acute diseases such as  pulmonary fibrosis and  ARDS.

\texttt{AngII} effects are mediated by two G-protein coupled receptors (GPCR) called  angiotensin II type 1 (\texttt{AT1R}) and type 2 (\texttt{AT2R}). In addition, it can be cleaved by different enzymes.  For example,  \texttt{ACE2}   generates   \texttt{Ang1-7} peptides and  aminopeptidase A (\texttt{APA}, EC 3.4. 11.7)  generates other peptides such as angiotensin III (\texttt{AngIII}) which is further cleaved to \texttt{AngIV}. In our model, we skipped all details about the  enzymatic reactions \texttt{AngII}-\texttt{AngIII}-\texttt{AngIV} and  kept only a single equation for their transformation.  The dynamics of \texttt{AngII} and \texttt{AngIV} can thus be  written as:

\begin{eqnarray}
 \frac{d[\texttt{AngII}]}{dt} &=&  \left( c_{ace} + c_{chy} \right) [\texttt{AngI}]\nonumber \\ &&- \left(c_{ace2} + c_{angIV} + c_{at1r}  + c_{at2r}\right)[\texttt{AngII}]- \frac{\texttt{Log}\; 2}{h_{angII}} [\texttt{AngII}]
     \end{eqnarray}

 \begin{equation}
 \frac{d[\texttt{AngIV}]}{dt} = c_{angIV} [\texttt{AngII}]- \frac{\texttt{Log}\; 2}{h_{angIV}} [\texttt{AngIV}]
     \end{equation}
\noindent     
where $h_{angII}$ and $h_{angIV}$ are the half-lives of the  peptides and $c_{ace2}$, $c_{angIV}$, $c_{at1r}$ and $c_{at2r}$ the rates of the  enzymatic reactions. 

The dynamics of the peptide-bound form of the GPCRs are expressed  as:

\begin{equation}
 \frac{d[\texttt{AT1R-AngII}]}{dt} = c_{at1r} [\texttt{AngII}] -\frac{\texttt{Log}\; 2}{h_{at1r}} [\texttt{AT1R-AngII}]
     \end{equation}

\begin{equation}
 \frac{d[\texttt{AT2R-AngII}]}{dt} = c_{at2r} [\texttt{AngII}] -\frac{\texttt{Log}\; 2}{h_{at2r}} [\texttt{AT2R-AngII}]
     \end{equation}
   
   \noindent  
 where [\texttt{AT1R-AngII}] and [\texttt{AT1R-AngII}] are the concentrations of the bound forms of the receptors, and $h_{at1r}$ and $h_{at2r}$
their half-lives.

Until now, we have modeled the \texttt{ACE}/\texttt{AngII}/\texttt{AT1R} regulatory axis of the RAS system. Since the last two decades, it became clear that there is another RAS axis that acts as a counterregulator of the  first axis 
\citep{REVIEWMAS}.
The key role of this second axis is played by the \texttt{Ang1-7} peptide
that is mainly produced from \texttt{AngII} by the \texttt{ACE2} enzyme and binds to the transmembrane GPCR called \texttt{MAS}. However, \texttt{Ang1-7} can  also be obtained as an enzymatic product from \texttt{AngI} via the catalytic activity of \texttt{NEP} and, to a lesser extent, from \texttt{Ang1-9} via \texttt{ACE} and \texttt{NEP}. We overlooked the \texttt{Ang1-9}-related enzymatic reactions in our model, as they contribute less to \texttt{Ang1-7} formation   \citep{Reninbook1,Reninbook2}. The dynamical equations for the \texttt{Ang1-7} peptide and  the \texttt{MAS}-bound receptor are as follows:

     \begin{eqnarray}
 \frac{d[\texttt{Ang1-7}]}{dt} = c_{nep}  [\texttt{AngI}]+ c_{ace2} [\texttt{AngII}]   - c_{mas} [\texttt{Ang1-7}]  - \frac{\texttt{Log}\; 2}{h_{ang1-7}} [\texttt{Ang1-7}]
 \label{Ang17}
     \end{eqnarray}
     
     \begin{equation}
 \frac{d[\texttt{MAS-Ang1-7}]}{dt} =  c_{mas} [\texttt{Ang1-7}]- \frac{\texttt{Log}\; 2}{h_{mas}} [\texttt{MAS-Ang1-7}]
 \label{MAS}
     \end{equation}

Let us now go back to Eq. (\ref{REN}) in which we simply expressed the renin production as a baseline term $\beta$. To describe the autoregulatory nature of the RAS system, this term has to depend on the production of other species, thus introducing a feedback regulation. It is known that this feedback depends on  \texttt{AT1R} bound to \texttt{AngII}. Following other models \citep{Leete18,Leete19}, we expressed $\beta$ as: 

 \begin{equation}
 \beta =  \beta_0 +  \left(\left(\frac{[\texttt{AT1R-AngII}]_0^N}{[\texttt{AT1R-AngII}]}\right)^{\delta}-1\right)\label{PPP}
 \label{feedback}
     \end{equation}
\noindent
where $\beta_0$ is a constant parameter to be identified and  $[\texttt{AT1R-AngII}]_0^N$  the equilibrium concentration  for healthy normotensive  humans. $\delta$ is a positive number that we fixed to 0.8  \citep{Leete18}.

Technical details on the procedure used to solve the model and on  model stability are given in the Materials and Methods section.

\subsection*{Modeling blood pressure and antihypertensive RAS-blocker effects}

Blood pressure is well known to be increased by the concentration of \texttt{AngII} bound to \texttt{AT1R}. It has also been described to be decreased by the concentration of \texttt{MAS} bound to \texttt{Ang1-7} and of \texttt{AT2R} bound to \texttt{AngII}, but the precise mechanism is not yet known \citep{MASDBP,MASBO,AT2DBP}. Therefore, we did not introduce in our model a feedback between these concentrations and renin production, as we did for \texttt{AT1R-AngII}, and  modeled the  blood pressure (DBP) simply from the \texttt{AT1R-AngII} concentration:
\begin{equation}
 DBP =  P_0 +  P_1 [\texttt{AT1R-AngII}]
 \label{DBP}
     \end{equation}
\noindent
We chose to fix the two parameters $P_0$ and $P_1$ to mimic the diastolic blood pressure (DBP)  rather than the systolic one. We thus fixed  DBP   equal to 80 mmHg for normotensive individuals and to 110 mmHg for hypertensive individuals. Hence, $P_0+P_1 [\texttt{AT1R-AngII}]_0^N =80$ mmHg and $P_0+ P_1 [\texttt{AT1R-AngII}]_0^H =110$ mmHg, where the $N$ and $H$  superscripts denote  concentration in normotensive and hypertensive individuals and the  $0$ subscript,   equilibrium concentrations. 


Since dysregulated  RAS with high levels of \texttt{AngII} are related to essential hypertension, a wide range of RAS-targeting drugs have been developed in the last fourty years \citep{DRUG01}. They can be classified in three different categories based on their pharamacological target \citep{DRUG02}:  

\begin{itemize}
\item Angiotensin-converting enzyme inhibitors (ACE-I) that bind to \texttt{ACE} and thus inhibit the formation of angiotensin II and the  associated vasoconstriction and inflammatory cascades. Examples of this type of drugs are  enalapril, lisinopril and  captopril.  

\item Angiotensin receptor blockers (ARB) that block the binding of \texttt{AngII} to  \texttt{AT1R} and thus act in antagonism with   \texttt{AngII}. Examples are  candesartan,  losartan and  valsartan.  

\item Direct renin inhibitors (DRI) that act on  renin and thus inihibit the conversion of \texttt{AGT} to \texttt{AngI}. Examples are aliskiren,  enalkiren and  remikiren. 
\end{itemize}

We modeled the action of these three types of drugs by modifying the reaction rates associated to their targets as:
\begin{eqnarray}
 c_{ace} \longrightarrow c_{ace} \times \left( 1 - \gamma_{\texttt{ACE-I}} \right)\nonumber\\
 c_{at1r} \longrightarrow c_{at1r} \times \left( 1 - \gamma_{\texttt{ARB}} \right)\nonumber\\
  c_{re} \longrightarrow c_{re} \times \left( 1 - \gamma_{\texttt{DRI}} \right)
  \label{gammadrugs}
\end{eqnarray}
\noindent
where $\gamma_{\texttt{ACE-I}}$, $\gamma_{\texttt{ARB}}$ and $\gamma_{\texttt{DRI}}$ are parameters describing the drug activity.  

\subsection*{Model predictions and clinical data on RAS-blocker drugs}

\begin{figure}[h!]
    \centering
    \includegraphics[width=0.97\linewidth]{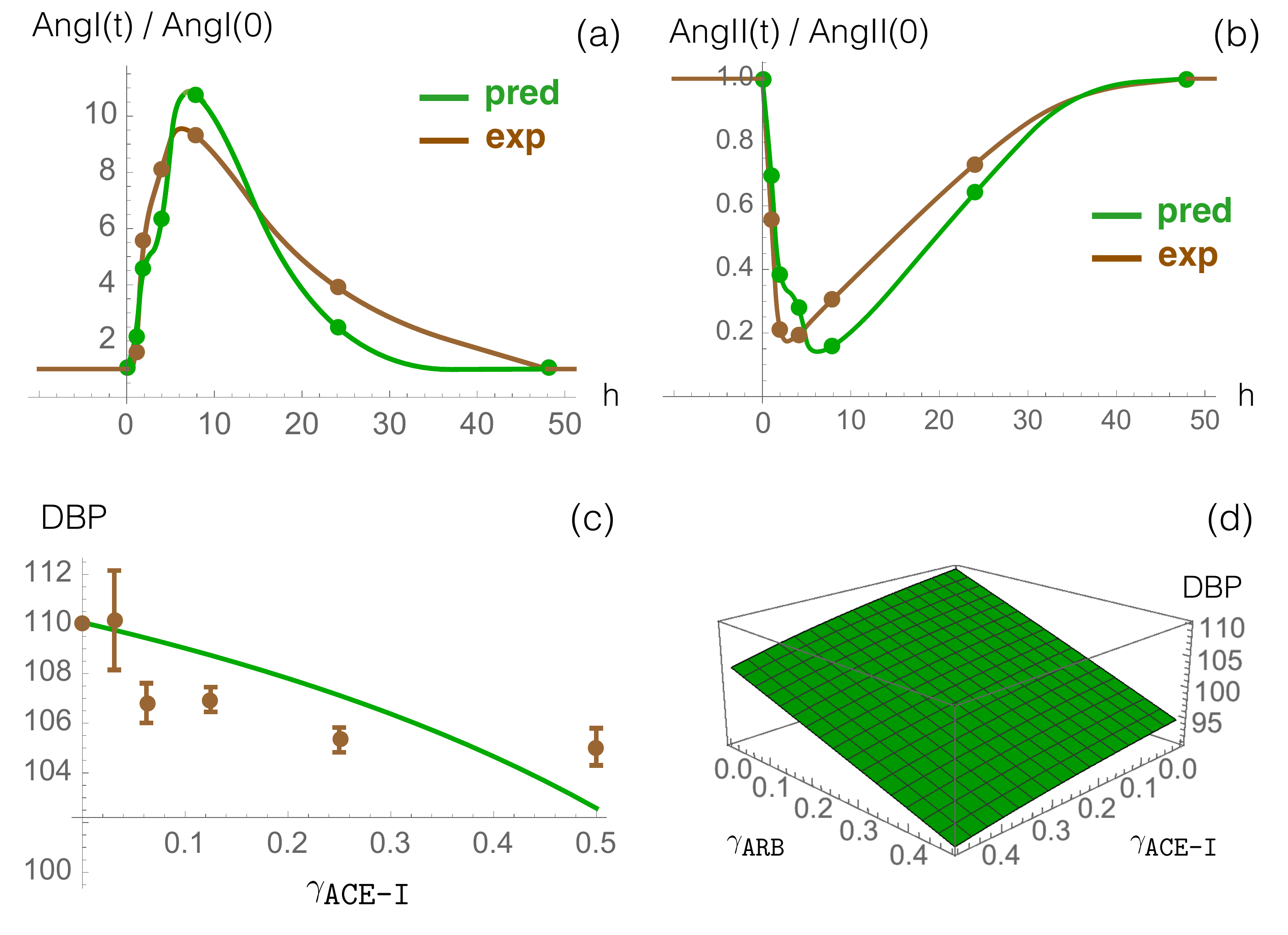}
    \caption{Dynamical response of RAS to  ACE-I (enalapril) administration. Comparison between the computational prediction (green) and the experimental data (brown) of  normalized \texttt{AngI} (a) and \texttt{AngII} (b) as a function of time (in hours) after the single dose administration. Continuum lines are obtained through data interpolation. (c) Predicted DBP as a function of $\gamma_\texttt{ACE-I}$ values (green line) and measured DBP  averaged  over more than ten ACE-I types as a function of the  dosage divided by the maximal dosage (brown points). (d) Predicted effect of the combination of ACE-I and ARB on the DPB values. }
    \label{1}
\end{figure}

The effect of enalapril, an ACE-I type drug, on plasma \texttt{ACE} activity and on plasma levels of \texttt{AngI} and \texttt{AngII}, has been measured in normotensive individuals who received a single oral dose of 20 mg  \citep{NUSS01}. To compare these data with model predictions, we first fitted the $\gamma_{\texttt{ACE-I}}$ parameter introduced in Eq. (\ref{gammadrugs}) to  the  \texttt{ACE} activity values during enalapril administration divided by the pre-treatment activity (measured by an antibody-trapping assay). Once $\gamma_{\texttt{ACE-I}}$ was set, we used our model to predict the dynamical response of RAS to this inhibitor drug. The time-dependent values of  \texttt{AngI} and \texttt{AngII} concentrations, normalized by their  concentration at time 0, are shown  in Figs \ref{1}.a-b both for our model predictions and experimental enalapril data; there is very good agreement between the two curves, without any further parameter fitting. The excellent correspondence between model prediction and experimental data is also clear from the root mean square deviation (rmsd) between model prediction and experimental data  on all time points following drug administration, as shown in Table \ref{POSSA}.

Our model, thus,  captures the known dynamics of \texttt{ACE} inhibition, (\emph{i.e.}, increased \texttt{AngI} levels and decreased  \texttt{AngII} levels);  this has the effect of lowering the concentration of \texttt{AngII} bound to \texttt{AT1R} and, thus, also lowers the blood pressure (Eq. (\ref{DBP})).

To study the effect of ARB antihypertensive drugs on RAS, we considered  data from \citep{NUSS02}, which measures the effects of different types of AT1R blocking molecules on plasma levels of \texttt{AngII} in normotensive individuals. Specifically, the study participants received a single 50 mg dose of losartan, 80 mg of valsartan or 150 mg of irbesartan. First we fitted the  $\gamma_{\texttt{ARB}}$ parameter (defined in Eq. (\ref{gammadrugs}))
to the \emph{in vitro} ability of the administered drug to induce the AngII receptor blockade, as measured by an AT1R radioreceptor binding assay \citep{NUSS02}. We then used our model to predict the time-dependent \texttt{AngI} level, which was normalized by its concentration prior to drug administration. The results were evaluated through the rmsd  between experimental and predicted values of \texttt{AngI}/\texttt{AngI}$_0$ at different time points after  drug administration. The results, which are detailed in Table \ref{POSSA}, clearly show that our model accurately predicts the RAS response to ARBs.

We also studied the effect of DRI-type drugs using experimental data that describe  \texttt{PRA} activity and  \texttt{RE}, \texttt{AngI} and \texttt{AngII} concentrations, when   
different doses of aliskiren were administered orally to normotensive individuals \citep{NUSS03}. 
We used the  \texttt{PRA} activity data to fit the  $\gamma_{\texttt{DRI}}$ parameter (introduced in Eq. (\ref{gammadrugs})) and we used our model to calculate the normalized \texttt{AngI} and \texttt{AngII} levels as a function of time. Here also, the results from our model and the experimental concentration data agree very well, as shown in Table \ref{POSSA}.

\begin{table}[h!]
\begin{center}
\begin{tabular}{ccccccc}
\hline
Drugs& Class& Dose & [\texttt{AngI}](t)/[\texttt{AngI}]$_0$  &  [\texttt{AngII}](t)/[\texttt{AngII}]$_0$  & Np &Ref.\\ 
 && (mg) & rmsd (range) &  rmsd (range) & & \\ \hline
Enalapril& ACE-I& 20 & 1.31 [1.0-9.2] &  0.09 [0.2-1.0] & 5 &  \citep{NUSS01} $\;$\\ 
Losartan &ARB & 50 & 0.61 [1.0-2.1] & - & 3 & \citep{NUSS02} $\;$ \\ 
Valsartan &ARB & 850 & 0.83 [1.0-2.2] &-  & 3 & \citep{NUSS02} $\;$ \\ 
Irbesartan& ARB & 150 & 0.97 [1.0-4.4] & - & 3 & \citep{NUSS02} $\;$\\
Aliskiren &DRI&   40  & 0.13 [0.4-1.1] & 0.14 [0.5-1.0] & 6 &  \citep{NUSS03} $\;$\\
Aliskiren& DRI& 80 & 0.15 [0.4-1.0] & 0.16  [0.4-1.0] & 6&\citep{NUSS03} $\;$\\ 
Aliskiren& DRI& 160 & 0.26 [0.2-1.0] & 0.20 [0.3-1.0]  & 6 &\citep{NUSS03} $\;$\\ 
Aliskiren& DRI& 640 & 0.29 [0.1-1.0] & 0.29 [0.1-1.0] & 6 & \citep{NUSS03} $\;$ \\ \hline
\multicolumn{3}{c}{\textbf{Mean}} & \textbf{0.57}\hspace{1.4cm}$\;$ & \textbf{0.18}\hspace{1.4cm}$\;$ &  & \\ \hline
\end{tabular}
\end{center}
\vspace{0.2cm}
\caption{Comparison between model predictions and experimental values of \texttt{AngI} and \texttt{AngII} levels normalized by their value before the administration of the drugs. Range is the interval of experimental values and rmsd is the root mean square deviation between experimental and predicted values, computed over all time points;   Np is the number of time points. }
\label{POSSA}
\end{table}

In summary, the rmsd  between predicted and experimental  values of normalized \texttt{AngI} and \texttt{AngII} levels, averaged over all  tested drugs, dosages, and a total of 38 time points, is 0.57 and 0.18, respectively (Table  \ref{POSSA}). These values  should be compared with average experimental values of 1.7 and 0.5 respectively, demonstrating excellent agreement between experimental data and model predictions.

It should be noted that all reported experimental data have been obtained after administration of single doses of RAS-targeting drugs. However, for hypertensive patients receiving long-term treatment, the expression of some enzymes involved in the RAS system could be either up- or down-regulated; we will return to this point in the Discussion section.

Finally, we compared model predictions against clinical data from large cohorts of patients describing the effect of ACE-I and ARB drug administration on blood pressure  \citep{LARGECOHORT2,LARGECOHORT1}. 
We first analyzed the response to ACE-I drugs alone. We plotted in Fig \ref{1}.c predicted DBP values as a function of $\gamma_{\texttt{ACE-I}}$, as well as measured DBP values  averaged over more than ten ACE-I drug types as a function of the normalized dosage \citep{LARGECOHORT2}.
For this, we fixed $\gamma_{\texttt{ACE-I}}=0.5$ at the maximum dosage and considered a linear relation between $\gamma_{\texttt{ACE-I}}$ and dosage. Note that it would have been possible to introduce additional parameters to define a non-linear relationship between these two quantities and, thus, obtain a better fit. Despite these simplifications, chosen to limit the number of parameters to fit, the model curve shows a reasonable fit to the experimental data.

 We then studied the effect of the combined administration of the two drugs, ARB and ACE-I, on blood pressure, plotting the predicted DBP values as a function of both $\gamma_{\texttt{ACE-I}}$  and $\gamma_{\texttt{ARB}}$ (see Fig. \ref{1}.d). We found that combined administration of ARB and ACE-I reduces DBP by 4 mmHg when compared with ARB monotherapy, and by 12 mmHg when compared with ACE-I monotherapy. These predictions should be compared with clinical DBP values of 3 mmHg for combined administration compared to either monotherapy \citep{LARGECOHORT1}.
 Thus, our model again provides an excellent prediction of experimental clinical data; further improvements to the model’s predictive strength might be possible by fixing the $\gamma_{\texttt{ARB}}$ value at the maximum dose to be slightly lower than the corresponding $\gamma_{\texttt{ACE-I}}$ value.

\subsection*{Modeling SARS-CoV-2 infection and ARDS severity}

\begin{figure}[h!]
    \centering
    \includegraphics[width=0.97\linewidth]{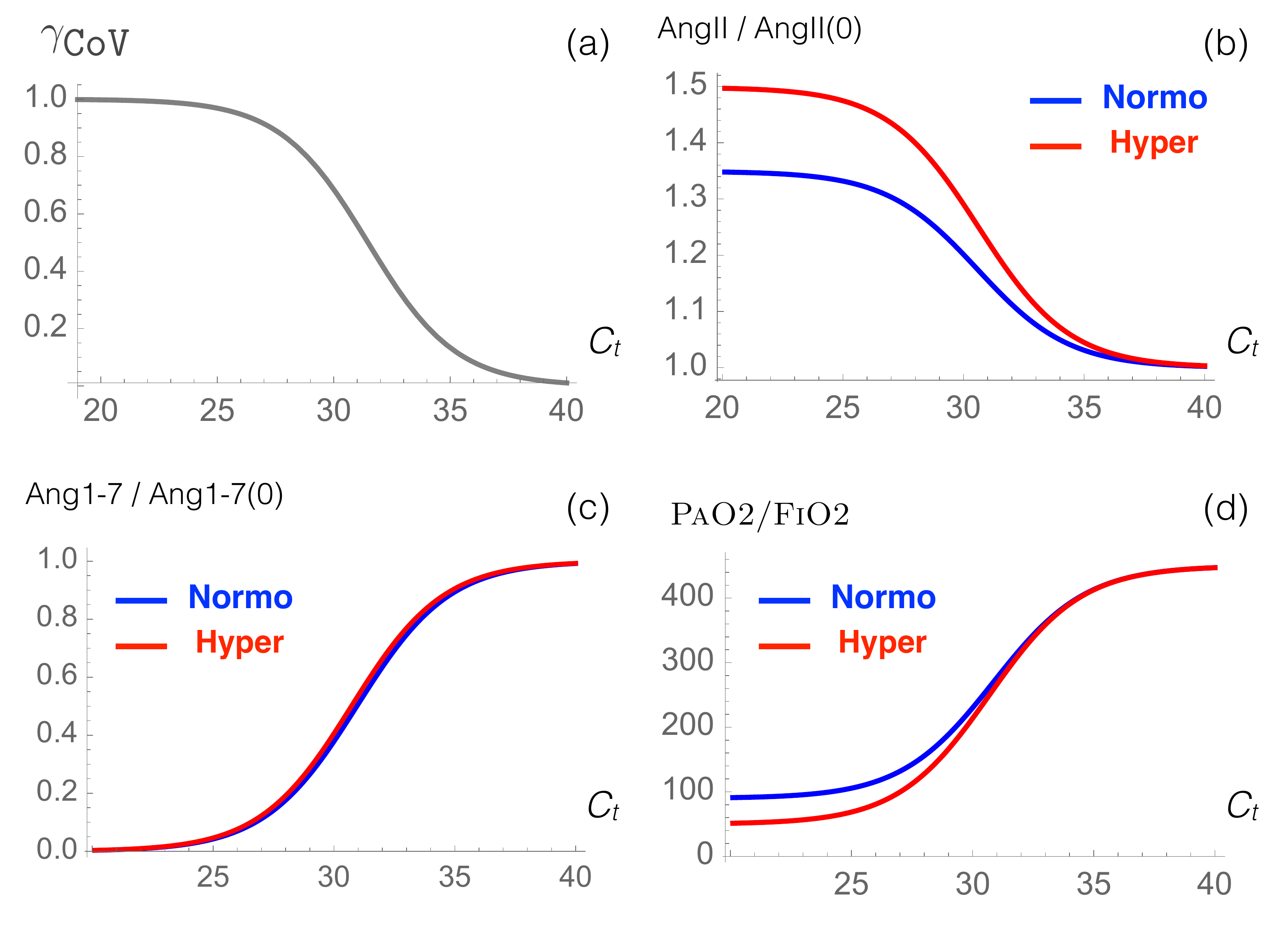}
    \caption{Simulated response of the RAS system to  viral infection. (a)  $\gamma_{\texttt{CoV}}$ function used to model the effect of the infection as a function of $C_t$, the cycle  threshold of the virus. (b)-(d) Predictions obtained from  our model for the  normalized levels of \texttt{AngII} and \texttt{Ang1-7}, and  for the physiological \textsc{PaO2/FiO2} value, as a function of $C_t$, for normotensive (blue) and hypertensive (red) individuals.}
    \label{1po}
\end{figure}

Since \texttt{ACE2} is the entry point of  SARS-CoV-2 \citep{WEN}, it is downregulated upon infection, and this impacts substantially on the local and systemic RAS systems. In order to model the downregulation effect due to the virus, we modified the \texttt{ACE2} reaction rate with the function $\gamma_{\texttt{CoV}}$:

\begin{equation}
 c_{ace2} \longrightarrow c_{ace2} \times \left( 1 - \gamma_{\texttt{CoV}}(C_t)\right)
\label{polli}
\end{equation}
\noindent
We chose $\gamma_{\texttt{CoV}}$ to be a sigmoid function of the cycle  threshold  value $C_t$, which is inversely related to the viral load \citep{ViralL}:

\begin{equation}
\gamma_{\texttt{CoV}}= \frac{1}{1+ e^{a C_t - b}}
\label{Ct}
\end{equation}
\noindent
where $a$ and $b$ are positive real numbers. $C_t$ values of 31.5, 27.6,  and  23.8 correspond  to mild, moderate and severe disease, respectively, and  $C_t > 40$ to undetected viral infection \citep{ViralL2}. We thus chose the inflection point of the sigmoid at $C_t= 31.5$ and imposed  $\gamma_{\texttt{CoV}}>0.99$ for $C_t > 40$. Using these relations, we identified the  values of the  parameters $a$ and $b$. They are reported in Table \ref{tablehalf}, and the sigmoid is represented in Fig. \ref{1po}.a.

To model ARDS severity and  how the lungs of SARS-CoV-2 infected patients evolve in response to RAS dysregulation, we introduced a phenomenological relation to estimate the \textsc{PaO2/FiO2} ratio, defined as  the ratio between the   partial pressure of arterial oxygen (\textsc{PaO2}) and the  fraction of inspired oxygen (\textsc{FiO2}). This quantity plays a key role in the assessment of ARDS patients \citep{ARDS,ARDS2}. The normal range of \textsc{PaO2/FiO2} is between 400 and 500 mmHg. Mild and moderate ARDS are characterized by \textsc{PaO2/FiO2}  values in the range [200–300] mmHg and [100-200] mmHg, respectively. ARDS is severe for values below 100 mmHg. 

We predicted the \textsc{PaO2/FiO2} ratio as a function of the \texttt{AngII} and \texttt{Ang1-7} concentrations:  

\begin{equation}
\textsc{PaO2/FiO2} = A_0+A_1 \left( -\frac{[\texttt{AngII}]}{[\texttt{AngII}]_0} +  \frac{[\texttt{Ang1-7}]}{[\texttt{Ang1-7}]_0}\right) 
\label{O2}
\end{equation}
\noindent
where $A_0$ and $A_1$ are two parameters that we identified on the basis of our model by comparing the baseline RAS  with the same system  in which \texttt{ACE2} is knocked out. In the former case we fixed \textsc{PaO2/FiO2}= 450 mmHg and in the latter \textsc{PaO2/FiO2}= 50 mmHg. 


\subsection*{RAS in COVID-19}

It is known that \texttt{ACE2} is the cellular receptor of the spike glycoprotein  of  SARS-CoV-2  \citep{ACE2,ACE2_0,ACE2_1,ACE2_2,ACE2_3}, and that it triggers the entry of  SARS-COV-2 into the host cell. Although \texttt{ACE2} is expressed in a variety of tissues \citep{ACE2EXP1,ACE2EXP2,ACE2EXP3}, it is expressed mainly in the alveolar epithelial cells of the lung, in the gastrointestinal tract and in the kidney proximal tubular cells. 

Here, we used our model to predict how the RAS system is perturbed by the SARS-CoV-2 virus. Simulation results for different \texttt{AngII} and \texttt{Ang1-7} concentrations, and for the physiological value of \textsc{PaO2/FiO2}, are presented in Figs 3.b-d and in Table \ref{TablePO}. 

We observe that the \texttt{AngII} level increases with increasing viral load, with a much stronger increase for hypertensive than for normotensive patients. The \texttt{AngII} level is predicted to increase by approximately 15\% for patients with moderate and severe COVID-19 (Table \ref{TablePO}); this prediction is in very good agreement with the experimental value of 16\% found in \citep{CLINICAL01}, but in poorer agreement with the  value of 35\% resulting from a study of only 12 patients \citep{CLINICAL02}. 

We also observe that our model predicts a severe reduction  of the \texttt{Ang1-7} level, due to \texttt{ACE2} downregulation; this reduction is the same for  hypertensive and normotensive patients.

The overall result of the model is that the RAS system becomes imbalanced upon SARS-CoV-2 infection, with  the harmful \texttt{AngII} axis upregulated and the counteracting \texttt{Ang1-7} axis  severely downregulated. This imbalance can be related to multiple clinical manifestations of COVID-19.
More specifically, increased \texttt{AngII} levels cause hyperinflammation which, in turn, increases plasma proinflammatory cytokine levels (in particular, IL-6) \citep{INFLAMMATION,INFLAMMATION02}.  In addition, thrombotic events are observed, since \texttt{AngII} promotes  the expression of plasminogen activator inhibitor-1 (PAI-1) and tissue-factors (TFs) \citep{TROMBOSI,TROMBOSII}.  \texttt{Ang1-7}, which normally counteracts  these various effects \citep{REVIEWMAS}, is downregulated by SARS-CoV-2 infection, such that the COVID-19  clinical manifestations become increasingly severe  as the  disease develops.

Moreover, our model predicts  severe ARDS with \textsc{PaO2/FiO2}$<$100 mmHg for normotensive and hypertensive patients whose $C_t$ values are smaller than 24.1 and 27.0, respectively. Our model predicts moderate ARDS, characterized by a \textsc{PaO2/FiO2} ratio in the range 100-200 mmHg, for normotensive  and hypertensive patients having $24.1<C_t<29.3$ and $27.0<C_t<29.7$, respectively, and  mild ARDS, characterized by a \textsc{PaO2/FiO2} ratio in the range 200-300 mmHg for normotensive  and hypertensive patients having $29.3<C_t<31.4$ and $29.7<C_t<31.6$, respectively.

Our modelling approach suggests a weak relationship between  hypertension and ARDS severity resulting from SARS-CoV-2 infection. The mean value of the $\textsc{PaO2/FiO2}$ ratio over the entire $C_t$ range is approximately 20 mmHg lower for hypertensive  than for normotensive patients. Indeed, the large difference in \texttt{AngII} levels between normotensive and hypertensive patients is partially compensated by the absence of any difference in  \texttt{Ang1-7} levels.

\begin{table}[]
\begin{center}
\begin{tabular}{ccccc}
\hline
 & Uninfected & Mild & Moderate  & Severe  \\
 $C_t$ & 40.0 & 31.5  & 27.6  & 23.8  \\ \hline
 \multicolumn{5}{c}{\hspace{2.5cm}Normotensive}   \\ \hline
$\;$[\texttt{AngII}] (fmol/ml) & 28 &32 & 36 & 38 \\
$\;$[\texttt{Ang1-7}] (fmol/ml)& 36 & 21 & 5 & 1 \\ 
\textsc{PaO2/FiO2} (mmHg)& 450 & 300 & 145  & 98 \\ 
DBP (mmHg)& 80 & 81  & 82  & 82  \\ \hline
\multicolumn{5}{c}{\hspace{2.5cm}Hypertensive}  \\ \hline
$\;$[\texttt{AngII}] (fmol/ml)& 156 & 186 & 221  & 231 \\ 
$\;$[\texttt{Ang1-7}] (fmol/ml)& 92 & 55 & 15 & 2 \\ 
\textsc{PaO2/FiO2} (mmHg)& 450 & 292 & 115 & 60  \\ 
DBP (mmHg)& 110 & 117  & 125  & 128 \\ \hline
\end{tabular}
\end{center}
\vspace{0.2cm}
\caption{Prediction of  biochemical and clinical features of SARS-CoV-2 infected patients.}
\label{TablePO}
\end{table}

\subsection*{Impact of RAS-modulating drugs on COVID-19 severity}

We analyzed the effect of administering a selection of drugs to normotensive and hypertensive patients who were infected with the SARS-CoV-2 virus. More specifically, we considered RAS-blocking drugs that are already commonly used to treat hypertension, as well as drugs that are currently undergoing clinical trials in the context of COVID-19, such as rhACE2 and Ang1-7. \newline

\begin{figure}[h!]
    \centering
    \includegraphics[width=0.975\linewidth]{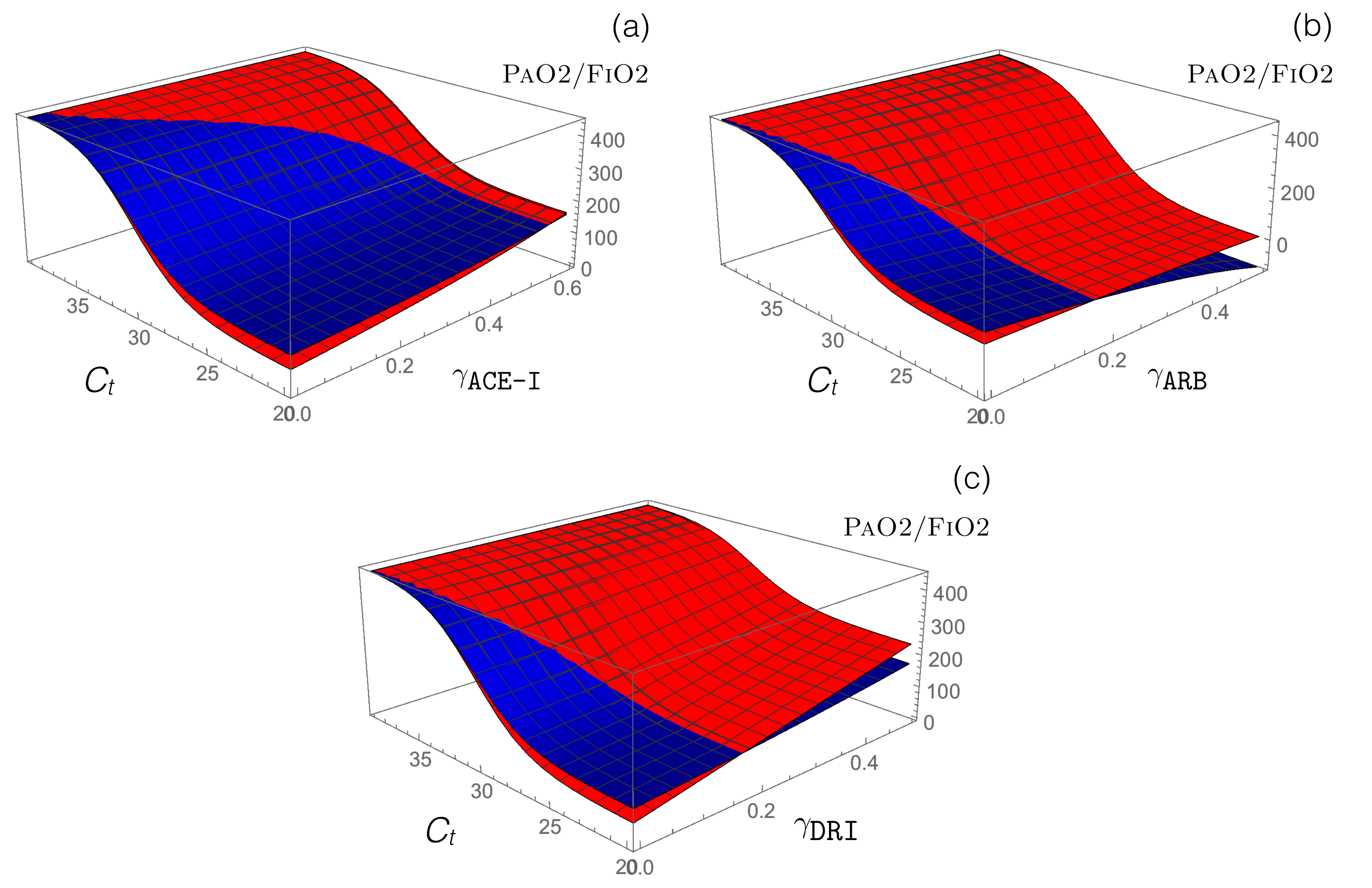}
    \caption{Impact of different RAS-blocking drugs in  normotensive (blue) and hypertensive (red) SARS-CoV-2 infected patients. Predicted $\textsc{PaO2/FiO2}$ value as a function of the cycle treshold value $C_t$ and (a) $\gamma_{ACE-I}$, (b) $\gamma_{ARB}$ and (c) $\gamma_{DRI}$ functions that model the administration of the corresponding drugs.}
    \label{DRUGCOVID}
\end{figure}

\noindent 
$\bullet$ \underline{Antihypertensive RAS-blocking drugs}. We combined the effect of each of the three RAS-blocking ACE-I, ARB and DRI drugs, which were modeled by the enzyme-inhibiting  $\gamma$ functions (introduced in Eq. (\ref{gammadrugs})), with the  \texttt{ACE2}-inhibiting $C_t$-dependent  $\gamma_{\texttt{CoV}}$ function (defined in Eq. (\ref{Ct})), which mimics SARS-CoV-2 infection. The \textsc{PaO2/FiO2} values predicted by our model are presented in Fig. \ref{DRUGCOVID}.

Our model predicts that administration of ACE-I and DRI drugs  protect from the adverse effects of ARDS, especially for hypertensive patients, while ARB drugs are predicted to worsen ARDS severity, especially for normotensive patients.

Model predictions for ACE inhibitors are in agreement with clinical data, which indicate that treatment with ACE inhibitors is associated with better survival among COVID-19 patients  \citep{BIGIF03,PERSOCONTO1}. Indeed, only 3\% of non-surviving COVID-19 patients that were monitored were treated with ACE-I drugs compared to 9\% of surviving COVID-19 patients \citep{BIGIF03}.
Moreover, in a meta-analysis \citep{PERSOCONTO1}, hypertensive patients treated with ACE-I drugs were associated with a reduced mortality of 35\% when compared to patients who were not treated with ACE-I drugs. In another clinical analysis \citep{PERSOCONTO2}, older patients who were treated with ACE-I drugs had a 40\% lower risk of hospitalization than those who were not treated with ACE-I drugs.

No data are currently available to validate our model prediction that COVID-19 disease attenuation due to ACE-I drug treatment is stronger in hypertensive than in normotensive patients. Furthermore, no data are currently available to validate our model prediction that DRI and ACE-I drug treatments cause similar levels of COVID-19 disease attenuation.

In contrast to DRI and ACE-I drugs, our model predicts that ARB drug treatment worsens COVID-19 disease severity, with the effect being stronger for normotensive compared to hypertensive patients. Here, the agreement between model predictions and clinical data is less clear, with some clinical data in  agreement with our model prediction \citep{BIGIF03,PERSOCONTO2}, while other clinical data suggest that ARB drug treatment does not affect hospitalization risk \citep{PERSOCONTO2} or mortality \citep{PERSOCONTO1,PERSOCONTO3}. This lack of agreement must be further investigated with additional clinical data.

Moreover, we performed a quantitative prediction of the drug effects on disease severity by calculating RAS peptide concentrations, \textsc{PaO2/FiO2} values and DPB of for moderate COVID-19 patients. Results are presented in Table \ref{tabledrugs}.

Administration of ACE-I drugs, modeled by $\gamma_{ACE-I}=0.5$, increases the $\textsc{PaO2/FiO2}$ value by approximately 50  and 70  mmHg for normotensive and hypertensive patients, respectively. An equivalent administration of  DRI drugs  increases this ratio even more, by 70 and 150 mmHg, while ARB administration decreases it by 140 and 30 mmHg for normotensive and hypertensive patients, respectively. 

The opposite effect of ARBs administration compared to ACE-I and DRI drugs can be attributed to the substantial increase in \texttt{AngII}  concentration, which is only partially balanced by a relatively small increase in \texttt{Ang1-7}  concentration, given that \texttt{ACE2} is downregulated in SARS-CoV-2 infection.

Note that a number of ARB drugs, including valsartan and losartan, are currently being tested in clinical trials, with the hope that they will rescue the RAS system  in COVID-19 patients  \citep{TRIAL4,TRIAL5,TRIAL6}. Our model predicts that this will not be the case.

Finally, as shown in Table \ref{tabledrugs}, the blood pressure is predicted to be unaffected by the administration of either ACE-I, ARB or DRI to normotensive COVID-19 patients, but to be reduced by approximately 10-20 mmHg by  administration  to hypertensive patients.\newline

\begin{table}[h!]
\begin{center}
\begin{tabular}{ccccccc}
\hline
\textbf{Drugs} & No Drugs & ACE-I   &  ARB & DRI  & rhACE2 & Ang1-7 \\ 
  \hline
\multicolumn{7}{c}{Normotensive - Moderate Infection}  \\ \hline
$\;$[\texttt{AngII}]/[\texttt{AngII}]$_{0}$& 1.29 & 1.10  & 1.98 &  0.99 &  1.10 & 1.29 \\ 
$\;$[\texttt{Ang1-7}]/[\texttt{Ang1-7}]$_{0}$& 0.15  & 0.13  & 0.23 & 0.11  & 0.68 & 0.64 \\ 
 \textsc{PaO2/FiO2} (mmHg) & 145 & 188 & 0 & 216 & 337 & 278  \\ 
  DBP  (mmHg) & 82  & 81 & 80 & 80 & 81 & 82  \\ \hline
\multicolumn{7}{c}{Hypertensive - Moderate Infection}  \\ \hline
$\;$[\texttt{AngII}]/[\texttt{AngII}]$_{0}$& 1.42  &  1.12 &  1.55 & 0.77 & 1.14 & 1.42 \\
$\;$[\texttt{Ang1-7}]/[\texttt{Ang1-7}]$_{0}$& 0.16  & 0.13  & 0.18 & 0.09 & 0.70 & 0.36  \\ 
 \textsc{PaO2/FiO2} (mmHg) & 115 &  185 & 83 & 268 & 332 & 167 \\ 
  DBP  (mmHg) & 125 & 114 & 101  & 102 & 115 & 125 \\ \hline
\end{tabular}
\vspace{0.2cm}
\caption{Predicted effects on \texttt{AngII} and \texttt{Ang1-7} levels, \textsc{PaO2/FiO2} and DBP upon drug administration by normotensive and hypertensive COVID-19 patients.  The drug administrations are modeled by $\gamma_{ACE-I}, \gamma_{\texttt{ARB}}, \gamma_{\texttt{DRI}}, \gamma_{rhACE2}=0.5$, $\eta_{Ang17} = 25$ fmol/(ml min) and moderate SARS-CoV-2 infection by $\gamma_{CoV}= 27.6$. }
\label{tabledrugs}
\end{center}
\end{table}

\noindent
$\bullet$ \underline{Other RAS-targeting drugs}. We used our model to test the potential of other drugs that are currently in clinical trials  to restore the functional activity of the perturbed RAS system upon viral infection. First, we modeled how the administration of an exogenous supplement of \texttt{rhACE2} (GSK2586881) affects  RAS  by modifying the  reaction rate $c_{ace2}$ defined in Eq. (\ref{polli}). This rate already includes the function $\gamma_{\texttt{CoV}}$ that mimics  SARS-CoV-2 infection, and we simply added a second function $\gamma_{\texttt{rhACE2}}$ associated with the effects of \texttt{rhACE2} administration: 

\begin{equation}
 c_{ace2} \longrightarrow c_{ace2} \times \left( 1 + \gamma_{\texttt{rhACE2}} - \gamma_{\texttt{CoV}}(C_t)\right)
\label{polli2}
\end{equation}
\noindent

\begin{figure}[h!]
    \centering
    \includegraphics[width=0.97\linewidth]{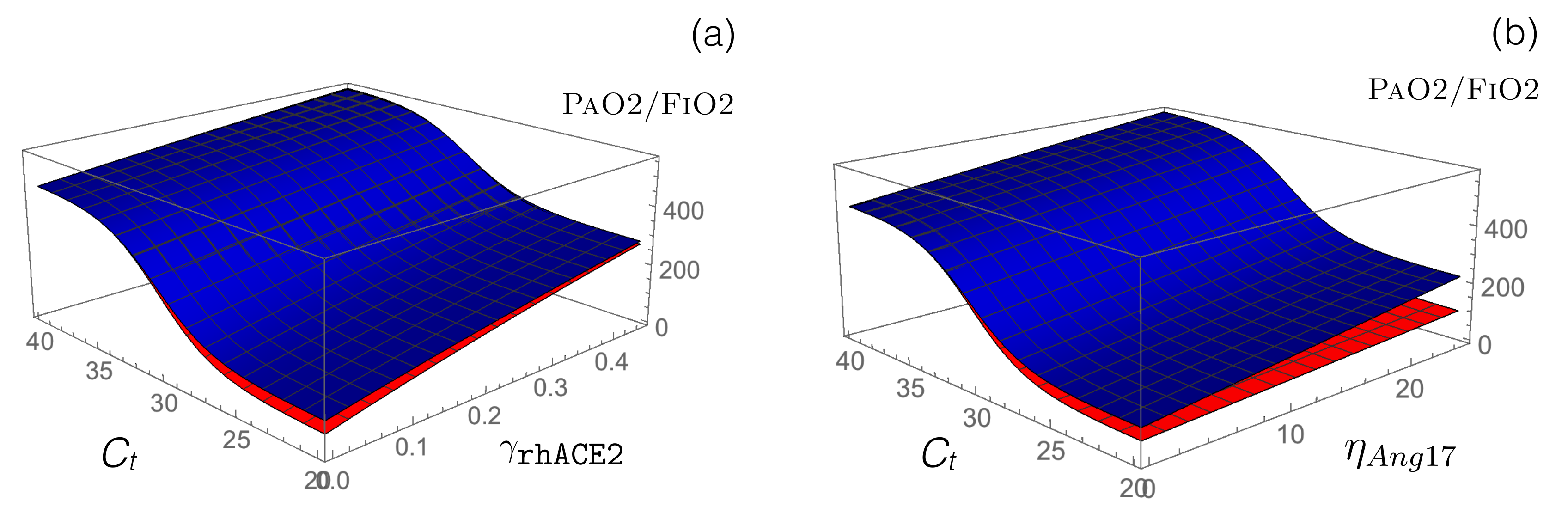}
    \caption{Impact on the $\textsc{PaO2/FiO2}$ value of the administration of \texttt{rhACE2} and \texttt{Ang1-7} in  normotensive (blue) and hypertensive (red) SARS-CoV-2 infected patients. (a) Predicted $\textsc{PaO2/FiO2}$ values as a function of $C_t$ and $\gamma_{\texttt{rhACE2}}$ (b) Predicted $\textsc{PaO2/FiO2}$ values as a function of $C_t$ and $\eta_{Ang17}$, the increase in the level of  \texttt{Ang1-7} due to its administration in infusion.}
    \label{777}
\end{figure}

Our model predicts an increase in $\textsc{PaO2/FiO2}$ following the administration of  exogenous \texttt{rhACE2}, thus predicting an alleviation of disease severity, as shown in Fig. \ref{777} and Table \ref{tabledrugs}. Specifically,  $\textsc{PaO2/FiO2}$ is predicted to increase by approximately 200 mmHg when $\gamma_{\texttt{rhACE2}}$ is fixed to 0.5. Our model also predicts, as expected, a reduction in \texttt{AngII} level and an increase in \texttt{Ang1-7} level.

These predictions are in agreement with both animal and \emph{in vitro} studies \citep{Imai2,Vanessa}, whereby  \texttt{rhACE2} is observed to alleviate virus-related ARDS severity through a double action. First, by \texttt{rhACE2} binding to the virus spike protein, interaction with endogenous \texttt{ACE2} is prevented and infection is slowed down.
Second, \texttt{rhACE2} administration increases \texttt{ACE2} activity, thus causing a reduction in \texttt{AngII} level and an increase in \texttt{Ang1-7} level; this protects the lung against severe failure. 

Current clinical trial data concerning the administration of different doses of \texttt{rhACE2} (0.1, 0.2, 0.4 and 0.8 mg/kg) to SARS-CoV infected patients
at different time intervals (2, 4, and 18 h), are only in partial agreement with our model predictions \citep{TRIAL1}.
Specifically, while clinical data followed the predicted decrease in  [\texttt{AngII}] and the predicted increase in [\texttt{Ang1-7}], there was no sustained increase in  $\textsc{PaO2/FiO2}$ compared with placebo. It has been suggested that the drug concentrations used in these clinical trials were too low to have a measurable effect on the respiratory system and that drug administration via infusion would have been more sustained  \citep{TRIAL1}. More experimental and clinical data are clearly needed to further investigate the effect of \texttt{rhACE2} on coronavirus-related ARDS. 

Another method of boosting the second RAS axis, \texttt{ACE2}/\texttt{Ang1-7}/\texttt{MAS},  which is downregulated by SARS-CoV-2 infection, is to administer   \texttt{Ang1-7} peptides as a means of triggering anti-inflammatory and anti-fibrotic mechanisms. We modeled \texttt{Ang1-7} peptide administration by introducing a new parameter, the production rate $\eta_{Ang17}$, to the dynamical Eq. (\ref{Ang17}) of [\texttt{Ang1-7}]; this allows the model to describe the exogenous \texttt{Ang1-7} level, which is added to the endogenous \texttt{Ang1-7} baseline. As shown in Fig. \ref{777}.b and  Table \ref{tabledrugs}, our model predicts a clear alleviation of COVID-19 severity, with $\textsc{PaO2/FiO2}$ increasing by 50 and 130 mmHg for hypertensive and normotensive patients, respectively,   upon administration of $\eta_{Ang17}=$ 25 fmol/(ml min) \texttt{Ang1-7} in infusion. Note that the COVID-19 alleviation is significantly stronger in normotensive compared to hypertensive patients for the same drug concentrations; a slightly stronger concentration of \texttt{Ang1-7} must be administered to hypertensive patients for an equivalent effect.

 Our model predicts a quantitative reduction in ARDS severity in COVID-19 patients, in agreement with the known anti-inflammation and anti-fibrosis nature of \texttt{Ang1-7}. Model predictions nicely agree with data from animal studies without the need of any additional fitting.
 For example, administration of \texttt{Ang1-7} by infusion to acid-injured rats suffering from ARDS increases baseline \texttt{Ang1-7} level by a factor 2.5, leading to an increase in $\textsc{PaO2/FiO2}$ of approximately 70 mmHg \citep{ICARE}. However, while the $\textsc{PaO2/FiO2}$ value increases linearly in our model as a function of \texttt{Ang1-7} concentration, it reaches a plateau in rats; this suggests that our model is probably oversimplified, since   $\textsc{PaO2/FiO2}$ is not a linear function of \texttt{Ang1-7} concentration. Further work on this aspect of our model will be possible when more data become available.

\section*{Discussion}

The spike protein of SARS-CoV-2 interferes with the RAS system by binding to the \texttt{ACE2} receptor, a key element of RAS. Despite recent progress in understanding the COVID-perturbed RAS system and how its functionality can be restored, more work is urgently needed in the context of the current COVID-19 pandemic.

We here present a simple computational approach to modeling RAS system evolution in the context of SARS-CoV-2 infection. Inspired by a number of existing RAS models \citep{Leete18,Leete19,Ford,Hallow14}, we  searched the literature for measured half-lives and concentrations of angiotensin peptides and their receptors  in healthy  normotensive  and hypertensive individuals, and then identified the unknown production and reaction rate parameters from the model.
As an initial test, we compared our model predictions of how the administration of RAS-blocking drugs would affect \texttt{Ang} peptide concentrations and blood pressure with relevant experimental data; we found good quantitative agreement between our model and experimental data, without the need for further parameter fitting. We then modeled the effect of SARS-CoV-2 infection on the RAS system through the downregulation of \texttt{ACE2}, which we related to the SARS-CoV-2 viral load.

A focal point of our work was to investigate how a series of RAS-targeting drugs affected
COVID-19 patients. We found that the administration of two antihypertensive drugs, ACE-I and DRI, tended to reduce the severity of COVID-19, while ARB drugs worsened it. Clinical data generally supports the model’s predictions for the administration of ACE-I drugs, but they are either absent or partially contradict the model’s prediction for DRI and ARB administration. 
Additionally, we modeled a potential treatment that is currently under clinical trial in COVID-19 patients: administration of \texttt{rhACE2} or \texttt{Ang1-7} by drug infusion. Our model predicts improved clinical outcomes in these cases, in agreement with a series of experimental data on animal models.

It is important to note that, despite its simplicity, our model has excellent accuracy in reproducing clinical and experimental data on the perturbed RAS system. Furthermore, the model’s predictions of changes in COVID-19 severity due to drug administration are blind predictions, without the fitting of any additional parameters.

Many challenges remain in our current understanding of RAS perturbation in COVID-19 patients. Importantly, more data regarding angiotensin peptide concentrations upon SARS-CoV-2 infection are urgently needed, since currently available data are often inconsistent or conflicting so that reliable comparisons between model predictions and experimental data cannot be  made. Even in healthy individuals, angiotensin peptide levels can vary substantially due to their low circulating concentrations, the experimental techniques used to measure them, and interpatient variability.

When developing our model we chose not to consider two enzymes that are active in the RAS system through the cancellation of their reaction rates: \texttt{CHY} and \texttt{NEP} (see Eqs (18)-(19)).
The \texttt{CHY}  enzyme is expressed in mast cells present in interstitial lung connective tissues, and it cleaves \texttt{AngI} to form \texttt{AngII}. The addition of this enzymatic reaction in the model would not really influence the predictions since it would essentially be    a reparametrization of \texttt{ACE} activity and of ACE-I action.  It might, nevertheless, be interesting to add the \texttt{CHY}  enzymatic reaction, which yields \texttt{ACE}-independent synthesis of \texttt{AngII} and  has been suggested (although debated) to be upregulated in the case of long-term ACE-I administration \citep{CHYM01}; this would enable an explanation of why ACE-I fails to inhibit \texttt{AngII} formation after some time \citep{CHYM01,CHYM02}. 

The \texttt{NEP} enzyme is expressed in a wide range of tissues, being particularly abundant in kidney, and it cleaves \texttt{AngI} to form \texttt{Ang1-7}. It  influences the counterregulatory RAS axis through its connection to \texttt{Ang1-7} levels,  thus affecting COVID-19 severity. However, \texttt{NEP}'s role is far from clear and the literature contains contradictory findings. Experimental data from rats with ARDS suggest that \texttt{NEP} is severely downregulated  in both plasma and  lung tissues \citep{NEP1}. Note that \texttt{NEP} also cleaves  natriuretic peptides, which have both anti-inflammatory and anti-fibrotic effects \citep{NEP}. Therefore, the combined administration of \texttt{NEP}-inhibiting and ARB drugs has been suggested to treat SARS-CoV-2 infected patients \citep{NEP2}.

Our future work will include building more complexity into our model by explicitly considering the communication between local and systemic RAS systems  \citep{Reninbook1,Reninbook2}, and by including the  interaction between  RAS and the immune system \citep{ImmuNEP}. This model extension  is necessary for an improved quantitative understanding of RAS system dysregulation upon a variety of perturbations, including SARS-CoV-2 infection.  

In summary, our model and its predictions provide a valuable and robust framework for \emph{in silico} testing of hypotheses regarding COVID-19 pathogenic mechanisms and the effect of drugs therapies that are aimed at restoring RAS functionality. Our work also opens a broader discussion on the role of the full RAS system in COVID-19, a topic that has received little attention to date, perhaps due to the current focus on the \texttt{ACE2} enzyme which, although very important as directly targeted by the virus, constitutes only one part of a much more complex system.

\section*{Materials and Methods}

\subsection*{Solving the RAS model}

The mathematical model of the RAS system described in Eqs (\ref{REN})-(\ref{feedback}) 
is a system of ordinary differential equations (ODEs), which are linear except for the
feedback loop of Eq. (\ref{feedback}). 

We collected from the literature  the values of the equilibrium concentrations of all proteins and peptides  except
renin and \texttt{MAS} bound 
to \texttt{Ang1-7}, for 
normotensive and hypertensive humans (Table \ref{tablecon}). From these values, we fixed the parameters that appear in the phenomenological relations (\ref{DBP}) and (\ref{O2}) for  DBP and \textsc{PaO2/FiO2} (Table \ref{tablehalf}).
We also collected the values of the half-life of all proteins and peptides but \texttt{MAS}; we assumed the latter to be equal to that of the other membrane receptors (Table \ref{tablehalf}). Moreover, we estimated the value of reaction rate $c_{re}$ from \citep{Ford,POLLO}.

Using these concentration and parameter values, we  solved the system of nine ODEs (Eqs (\ref{REN}) and (\ref{AGT})-(\ref{MAS})) at the stationary state to identify the unknown parameters and concentrations. However, these equations have 12 unknowns: $k_{agt}$, $\beta_0$, $c_{ace}$, $c_{ace2}$, $c_{angIV}$, $c_{at1r}$, $c_{at2r}$, $c_{mas}$, $c_{chy}$, $c_{nep}$, [\texttt{RE}]$_0$ and [\texttt{MAS-Ang1-7}]$_0$.
We had thus to assume  three additional relations, which are:
\begin{eqnarray}
c_{mas}&=& c_{at2r}\\
c_{chy}&=&0 \label{chy}\\
c_{nep}&=&0 \label{nep}
\end{eqnarray}
\noindent
Since no quantitative data related to the \texttt{MAS} receptor can be found in the literature, we hypothesized the first relation assuming  \texttt{MAS} and \texttt{AT2R} to be equally expressed and  the affinity of \texttt{Ang1-7} for \texttt{MAS} to be similar to the affinity of  \texttt{AngII} for \texttt{AT2R} \citep{MASBO}.  Moreover, we assumed $c_{chy}=0$ and  $c_{nep}=0$,  but   discussed the effect of non-vanishing values in the Discussion section. 

By imposing these three additional relations, we solved the system of 9 ODEs  at the stationary state. The values obtained for [\texttt{RE}]$_0$ and [\texttt{MAS-Ang1-7}]$_0$, $k_{agt}$, $\beta_0$, $c_{ace}$, $c_{ace2}$, $c_{angIV}$, $c_{at1r}$ and $c_{at2r}$ for normotensive and  hypertensive humans are given in Table \ref{tablecon}.

\vspace{0.5cm}
\begin{table}[h!]
\begin{center}
\begin{tabular}{ccccc}
\hline
Parameter & Unit & Normotensive &  Hypertensive  & Reference \\ \hline
$\;$[\texttt{AGT}]$_0$ & fmol/ml & 6 $\times 10^5$ & 6 $\times  10^5$ & \citep{Katsurada07}  \\
$\;$[\texttt{AngI}]$_0$ &  fmol/ml & 70  & 110  & \citep{Chappel16} \\
$\;$ &   &   &  & \citep{Pender} \\
$\;$[\texttt{AngII}]$_0$ &  fmol/ml & 28  & 156 & \citep{Chappel16}\\
$\;$ &   &   &  & \citep{Pender} \\
$\;$[\texttt{Ang1-7}]$_0$  &fmol/ml & 36  & 92 &  \citep{Chappel16} $\;$\\
$\;$ &   &   &  & \citep{Pender} \\
$\;$ &   &   &  & \citep{Sullivan15} \\
$\;$[\texttt{AngIV}]$_0$  & fmol/ml & 1 & 1 & \citep{Nussemberg86}\\
$\;$[\texttt{AT1R-AngII}]$_0$  &fmol/ml & 15 & 85 &  \citep{Leete18} \\
$\;$[\texttt{AT2R-AngII}]$_0$  &fmol/ml & 5 & 27 &    \citep{Leete18} \\ 
\hline
$\;$[\texttt{RE}]$_0$  &fmol/ml & 9.43 & 25.25 & Solved\\
$\;$[\texttt{MAS-Ang1-7}]$_0$  &fmol/ml & 6.43 & 15.92 &    Solved\\
$k_{agt}$  &fmol/(ml min) & 881.82 & 1198.22 &    Solved \\
$\beta_0$  &fmol/(ml min)& 0.54 &  2.21 &    Solved \\
$c_{ace}$  &1/min & 1.31 & 3.21 &    Solved \\ 
$c_{ace2}$  &1/min  & 1.80 & 0.82 &    Solved \\ 
$c_{angIV}$  &1/min  & 0.05 & 0.01 &    Solved \\
$c_{at1r}$  &1/min  & 0.03 & 0.03 &    Solved \\
$c_{at2r}$  &1/min  & 0.01 & 0.01 &    Solved \\ 
\hline
\end{tabular}
\vspace{0.2cm}
\caption{Equilibrium concentrations of the species involved in  RAS,  and production and reaction rate parameters, for healthy normotensive and hypertensive humans. 'Solved' means solved from the model.}
\label{tablecon}
\end{center}
\end{table}

\begin{table}[ht!]
\begin{center}
\begin{tabular}{cccc}
\hline
Parameter & Unit & Values  & Reference \\ \hline
 $h_{agt}$  & min & 600 & \citep{Hallow14}  \\
 $h_{ang1-7}$  & min & 0.5 & \citep{Hallow14}  \\ 
 $h_{angI}$  & min & 0.5 &  \citep{Hallow14} 
 \\
 $h_{angII}$  & min & 0.5  &\citep{Hallow14}
 \\ 
$h_{angIV}$  & min & 0.5 & \citep{Hallow14} 
 \\ 
$h_{at1r}$  & min & 12 & \citep{Hallow14}
 \\ 
$h_{at2r}$  & min & 12 &  \citep{Hallow14}  \\ 
$h_{re}$  & min & 12 &  \citep{Hallow14}  \\ 
$h_{mas}$  & min & 12 &   - \\ 
$c_{re}$  & 1/min & 20 &   \citep{Ford} \\ 
 & & &   \citep{POLLO} \\ 
\hline
$A_0$ & mmHg & 450  & Fitted \\
$A_1$ & mmHg & 267 & Fitted \\ 
$P_0$ & mmHg & 73.6 & Fitted \\
$P_1$ & mmHg ml/fmol & 0.43 & Fitted \\ 
$a$ & - & 0.53 & Fitted \\ 
$b$ & - & 16.7 & Fitted \\ 
\hline
\end{tabular}
\end{center}
\vspace{0.2cm}
\caption{Half-lives of the   species involved in  RAS   and other parameters of the model. 'Fitted' means fitted on experimental data. }
\label{tablehalf}
\end{table}

\subsection*{Stability of the RAS model}

The system of nine ODEs (Eqs (\ref{REN}) and (\ref{AGT})-(\ref{MAS})) can be summarized in the form:

\begin{equation}
 \frac{dx(t)}{dt} = f(x(t),\theta)	
\end{equation}
where $x(t)$ is the vector containing the nine state variables, \emph{i.e.} the concentrations of all proteins and peptides at time $t$,  $\theta$ is the vector with all the production, kinetic  and half-live parameters, and $f$ represents the vector that corresponds to the right-hand sides of Eqs (\ref{REN}) and (\ref{AGT})-(\ref{MAS}). 
In order to analyze the stability of the two steady states $x_0^N$ and $x_0^H$ for normotensive and hypertensive individuals, respectively,   we computed the eigenvalues of the Jacobian matrix:
\begin{equation}
 J(x_0) = \frac{\partial f(x,\theta)}{\partial x}\Bigr|_{x=x_0}	
\end{equation}
\noindent
where $x_0$ stands  for either $x_0^N$ or $x_0^H$. 

In both the normotensive and hypertensive cases, seven strictly negative real values were obtained, together with two complex conjugate eigenvalues  with strictly negative real parts. Both steady-states $x_0^N$ and $x_0^H$ are therefore stable. The nonzero imaginary parts of the two complex conjugate eigenvalues are responsible of some damped oscillations in transient responses to parameter changes, but the  overshoots are  limited. 
It is interesting to note that the imaginary part is more than three times lower in the hypertensive case, hence leading to more damped responses in comparison with the normotensive case. 

To quantify the state variable transients and the  aforementioned overshoots, we simulated step responses corresponding  to 10\% increase in the normal baseline for renin production $\beta_0$.  We  observe some damped oscillations during the transient phase of the normotensive case,  with very limited overshoots, \emph{e.g.} 1.3\% for  \texttt{RE} concentration. In the hypertensive case, the imaginary part of the complex conjugate eigenvalues is so low that the overshoots become almost undetectable (0.025\%).

\subsection*{Data availability}
The code used to generate all the results of this paper freely is available on GitHub\\ (https://github.com/3BioCompBio/RASinCOVID).

\subsection*{Acknowledgements}
We thank Dr. Filippo Annoni and Prof. Fabio Taccone for enlightening discussions. FP and MR are Scientific Collaborator and Research Director, respectively, at the  F.R.S.-FNRS Fund for Scientific Research.

\subsection*{Conflict of Interest}
The authors declare that they have no conflict of interest.

\bibliographystyle{apalike} 
\bibliography{msb}

\end{document}